\documentclass{webofc}

\usepackage[varg]{txfonts}   
\usepackage{hyperref}
\usepackage{url}
\hypersetup{colorlinks=true,citecolor=blue,urlcolor=blue,linkcolor=blue}
\begin{document}
\title{High-precision determination of radiative corrections to superallowed nuclear beta decays}
%
%

\author{\firstname{Chien-Yeah} \lastname{Seng}\inst{1,2,3}\fnsep\thanks{\email{cseng@utk.edu}}
}

\institute{Facility for Rare Isotope Beams, Michigan State University, East Lansing, MI 48824, USA 
\and
           Department of Physics, University of Washington,
           Seattle, WA 98195-1560, USA
\and
           Department of Physics and Astronomy, University of Tennessee, Knoxville, Tennessee 37996, USA
          }

\abstract{Superallowed $0^+\rightarrow 0^+$ transitions between $T=1$ nuclei have been a perfect avenue avenue for determining the Cabibbo-Kobayashi-Maskawa matrix element $V_{ud}$, which imposes powerful constraints on physics beyond the Standard Model at low energies. For a long time, the precision of $V_{ud}$ has been limited by uncertainties in radiative corrections that arise from non-perturbative strong interaction physics at both the hadronic and nuclear levels. In this talk, I will describe some recent efforts to pin down these corrections by combining dispersive analysis with experimental data, lattice QCD, and nuclear many-body calculations.
}

\maketitle
\section{Introduction}
\label{intro}

This is an invited talk presented in the 14th International Spring Seminar on Nuclear Physics: ``Cutting-edge developments in nuclear structure physics''. 

Despite being one of the most successful theories in physics ever, the Standard Model (SM) of particle physics is believed to be an incomplete theory due to its inability to answer many fundamental questions in physics, such as the existence of dark matter, dark energy, the matter-antimatter imbalance, and the nature and origin of neutrino masses. Beta decays, namely weak decays of strong interaction bound states through the emission of a W-boson which turns into an $e\bar\nu_e$ (or $e^+\nu_e$) pair, provide a perfect avenue to search for physics beyond the Standard Model (BSM) at low energies. By measuring observables in the decay process to extremely high precision and comparing them to SM theory predictions at the same level of precision, one may search for deviations between theory and experiment which indicate the existence of BSM physics. 

\begin{figure}[tb]
	\centering
	\includegraphics[scale=0.5]{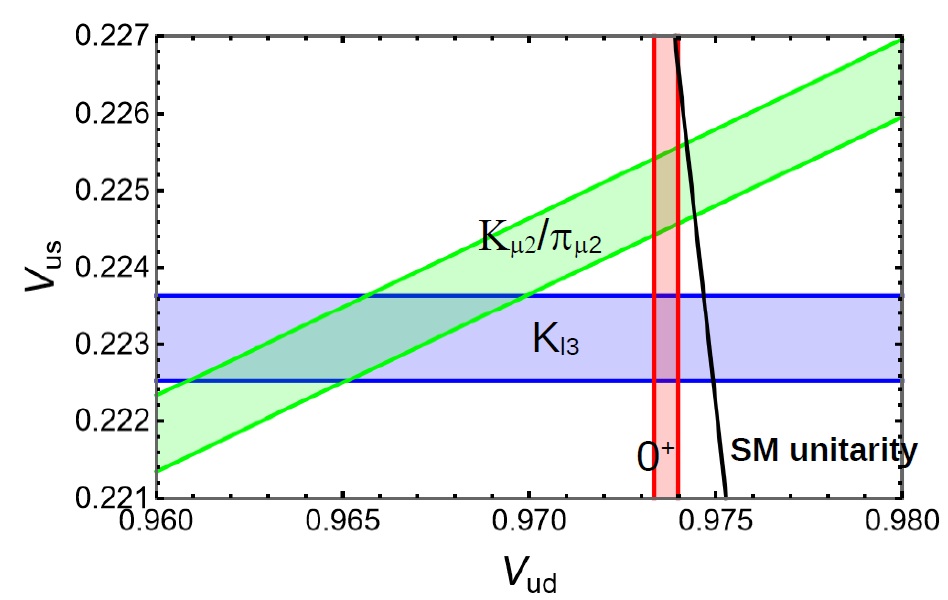}\hfill
	\caption{\label{fig:VudVus} Summary of the best-measured values of $V_{ud}$ (from superallowed $0^+\rightarrow 0^+$ decays) and $V_{us}$ (from leptonic decays of kaons and pions, $K_{\mu 2}/\pi_{\mu 2}$, and from semileptonic decays of kaons, $K_{\ell 3}$), in comparison to the SM unitarity (black line). The non-existence of a common overlapping region between the color bands and the black line indicates a violation of the unitarity relation, Eq.\eqref{eq:unitarity}.} 
\end{figure}

One of such quantities that can be precisely measured in beta decays is $V_{ud}$, the upper-left element of the Cabibbo-Kobayashi-Maskawa (CKM) matrix~\cite{Cabibbo:1963yz,Kobayashi:1973fv} that describes the mixing of quark weak flavor eigenstates. It is a unitary matrix according to the SM, which imposes specific constraints on its matrix elements. For instance, the first-row matrix elements have to satisfy the following unitarity relation:
\begin{equation}
	|V_{ud}|^2+|V_{us}|^2+|V_{ub}|^2=1~.\label{eq:unitarity}
\end{equation}
This relation is currently being tested to $\sim$0.01\% level, which can be translated into constraints on new physics at multi-TeV scale, competitive to high-energy experiments at colliders. This field has recently drawn substantial attentions from the physics community, as an apparent violation of Eq.\eqref{eq:unitarity} at $\sim 3\sigma$ level is observed based on th most precisely measured values of $V_{ud}$ and $V_{us}$, best summarized in Fig.\ref{fig:VudVus}. 

At the present, superallowed $0^+\rightarrow 0^+$ beta decays of $T=1$ nuclei provide the best determination of $V_{ud}$~\cite{Hardy:2020qwl}. This decay mode possesses a few obvious advantages comparing to other channels such as free neutron decay and pion decay: (1) It is a pure Fermi transition, which means the tree-level decay amplitude is completely fixed by isospin symmetry, barring small corrections from isospin symmetry breaking (ISB) effects; (2) There are 23 measured superallowed nuclear transitions, 15 of which lifetime precision is better than 0.23\%. Averaging over all such transitions results in a substantial reduction of experimental uncertainties. As a result, the largest source of uncertainties in the $V_{ud}$ extraction from superallowed decays comes not from experiment, but rather from the theory uncertainties in the calculation of the higher-order SM corrections to the decay lifetime. Therefore, future breakthroughs along this direction inevitably requires an improved theory prediction of such corrections within the SM.

\section{Overview of radiative corrections}

\begin{figure}[tb]
	\centering
	\includegraphics[scale=0.55]{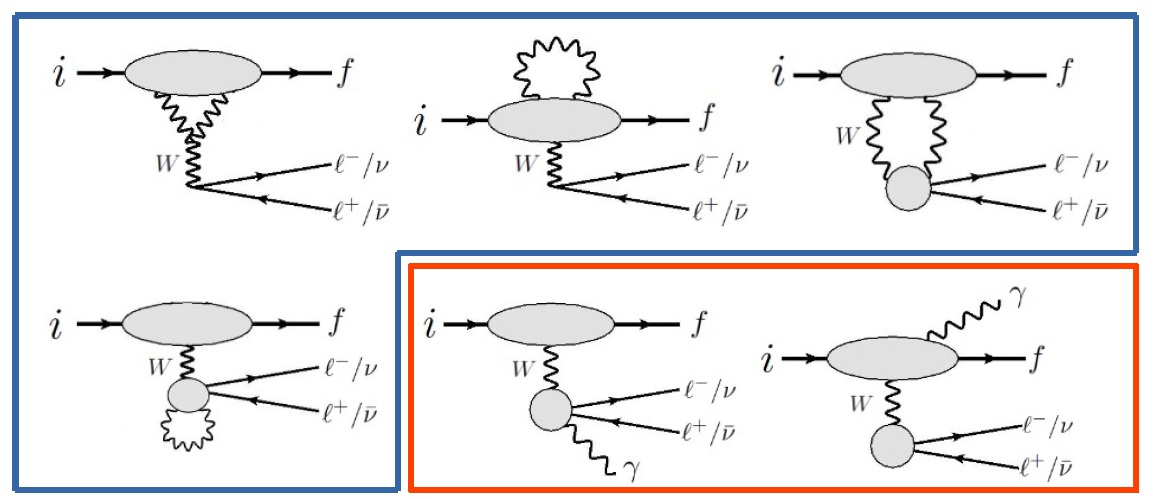}\hfill
	\caption{\label{fig:EWRC} Feynman diagrams for $\mathcal{O}(\alpha/\pi)$ electroweak RC to superallowed beta decays, with $i$ and $f$ denoting the initial and final nucleus. Diagrams in blue and red boxes represent virtual and real (bremsstrahlung) corrections, respectively.} 
\end{figure}

In this talk I will focus on one such important SM effects, the electroweak radiative corrections (RC) to the superallowed transition rate. They represent perturbations induced by the emission and reabsorption of virtual gauge bosons (loop corrections) as well as the emission of real photons (bremsstrahlung), see Fig.\ref{fig:EWRC}. A typical size of such corrections is $\alpha/\pi\sim 10^{-3}$, but they can be enhanced, for example, by large logarithms. So, to extract $V_{ud}$ to a  $10^{-4}$ level requires one to compute the RC with high accuracy. 

It is customary to split the full RC to the nuclear beta decay into two pieces:
\begin{equation}
	\text{Nucleus RC = (Nucleon RC) + (Nucleus RC - Nucleon RC)}
\end{equation}
The first term at the right hand side gives the quantity $\Delta_R^V$ which represents the RC to the free neutron decays, while the second term represent the ``difference'' between the nuclear and nucleon RC and is denoted as $\delta_\text{NS}$. 

\subsection{Nucleus-independent RC}

The nucleus-independent RC $\Delta_R^V$ appears in the decay of free neutron. Among all the $\mathcal{O}(\alpha/\pi)$ contributions, the main source of theory uncertainty stems from the so-called $\gamma W$-box diagram, i.e. the third diagram in the first row of Fig.\ref{fig:EWRC}, with the second gauge boson being a photon. Its precise determination is challenging because the virtual photon probes the loop momentum $q$ at all scales; at $q\sim 1$~GeV, the shaded blob on the nucleon side is governed by Quantum Chromodynamics (QCD) in its non-perturbative regime, which makes the precise calculations very challenging.

There are currently two reliable methods to compute the single-nucleon $\gamma W$-box diagram. The first is the dispersion relation (DR) analysis introduced in Refs.\cite{Seng:2018yzq,Seng:2018qru}. With Cauchy's theorem and optical theorem, one may relate the $q$-integration of the shaded blob to that of a parity-odd single nucleon structure function $F_3(x,Q^2)$, where $x$ is the Bjorken variable and $Q^2=-q^2$. The most theoretically-uncertain part of the structure function, which resides in the small-$x$, small-$Q^2$ regime, is governed by Regge physics and can be inferred from inclusive neutrino-nucleus scattering. Averaging over the results of several DR-based analysis returns $\Delta_R^V=0.02479(21)$~\cite{Gorchtein:2023srs}.  

The second method is to directly compute the box diagram using lattice QCD. It is done by calculating the upper shaded blob in the $\gamma W$-box diagram as a ``four-point correlation function'': 
\begin{equation}\int d^4xe^{iq\cdot x}\langle p|T\{J_\text{em}^\mu(x)J_W^\nu(0)\}|n\rangle~,
\end{equation}
which involves a time-ordered product of the electromagnetic (em) current and the charged weak (w) current. The combination of the lattice calculation of this function at $Q^2<2$~GeV$^2$ and the perturbative QCD calculation at $Q^2>2$~GeV$^2$ fully pins down the box diagram integrand from first-principles. This program was first carried out for the pion box diagram~\cite{Feng:2020zdc} and later for the nucleon box diagram~\cite{Ma:2023kfr}. The latter, in particular, reports $\Delta_R^V=0.02439(18)$ which agrees well with the DR result within 1.5$\sigma$. The slightly lower central value of the lattice determination might stem from the elastic (Born) contribution to the box diagram, which is a topic for future investigations. 

\subsection{Nucleus-dependent RC: dispersive approach}

Next, I will describe the the nucleus-dependent RC $\delta_\text{NS}$. In the DR language, it originates from the difference between the parity-odd structure function $F_3$ at the nucleon and nuclear level. At low energies, it is contributed by the nuclear resonance peaks and the Fermi-broadening of the single-nucleon Born contribution (the so-called quasi-elastic peak); at high energies, it is contributed by the nuclear modification to the single-nucleon structure function, including the nuclear shadowing, anti-shadowing and the EMC effect. 

$\delta_\text{NS}$ was first computed with traditional shell model~\cite{Barker:1991tw,Towner:1992xm,Towner:1994mw,Towner:2002rg,Towner:2007np}. Such computations, however, were shown to have missed important nuclear effects, such as the contribution from the quasi-elastic absorption peak~\cite{Seng:2018qru}. 
The DR representation provides an appropriate framework to study $\delta_\text{NS}$ from first principles, through the computation of the low-energy nuclear absorption spectrum with nuclear \textit{ab initio} methods~\cite{Seng:2022cnq,Gorchtein:2023naa}. There are two equivalent ways to state the problem:
\begin{eqnarray}
		\text{RC Integrand} &\sim & \langle f|J_\text{em}(\vec{q})G(M+q_0+i\varepsilon)J_W(-\vec{q})+\dots|i\rangle\nonumber\\
		&\sim& \sum_X\delta(q_0+M-E_X)\langle f|J_\text{em}(\vec{q})|X\rangle \langle X|J_W(-\vec{q})|i\rangle~,
\end{eqnarray}
namely the integrand of the nuclear $\gamma W$-box diagram integral can be expressed either in terms of the nuclear Green's function $G(z)=1/(z-H_\text{QCD})$ (first line), or the nuclear response function (second line). 

\begin{figure}[tb]
	\centering
	\includegraphics[scale=0.55]{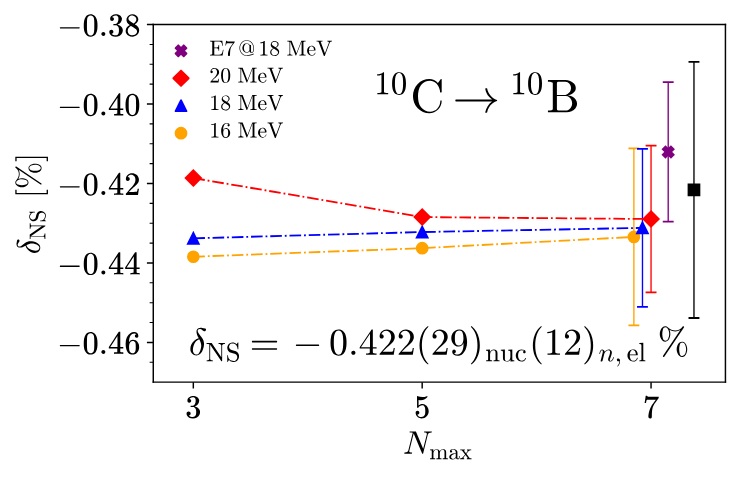}\hfill
	\caption{\label{fig:Nmax} The convergence of the NCSM calculation of $\delta_\text{NS}$ in ${}^{10}\text{C}\rightarrow {}^{10}\text{B}$ with increasing $N_\text{max}$. Lines with different colors represent calculations with different oscillator frequencies $\omega$. Figure taken from Ref.\cite{Gennari:2024sbn}. } 
\end{figure}

The first \textit{ab initio} calculation of $\delta_\text{NS}$ based on the method outline above is performed in Ref.\cite{Gennari:2024sbn} for the lightest measured superallowed transition, ${}^{10}\text{C}\rightarrow {}^{10}\text{B}$. The adopted many-body method is the No Core Shell Model (NCSM)~\cite{Barrett:2013nh}, which consists of expanding the nucleon wavefunctions in harmonic oscillator basis up to a given maximum principal quantum number $N_\text{max}$; two- and three-body chiral forces fitted from experiments are adopted for the nuclear interaction. The Lanczos strength method is used to simplify the summation over the nuclear intermediate states in the computation of the nuclear Green's function~\cite{haydock1974inverse,Dagotto:1993ajt,Marchisio:2002jx}, while the electroweak currents are expanded in terms of multipole operators~\cite{Donnelly:1975ze,Donnelly:2017aaa}. A rapid convergence of the result with respect to increasing $N_\text{max}$ is observed, see Fig.\ref{fig:Nmax}.

It is important to notice that, such calculations are based on low-energy chiral interactions with nucleons as the only degree of freedom. As a result, they are intrinsically incapable to account for contributions to $\delta_\text{NS}$ from non-nucleonic degrees of freedom; this missing piece must be included as a systematic uncertainty in the final result. In Ref.\cite{Gennari:2024sbn}, this uncertainty was estimated using the experimental data of nuclear shadowing correction to the parity-even nucleon structure function $F_2$. This na\"{\i}ve treatment can be improved in the future by explicitly computing the shadowing correction to $F_3$ using, for example, the Glauber-Gribov method~\cite{glauber1959,Gribov:1968jf}.

It is instructive to review the ``evolution'' for $\delta_\text{NS}$ for ${}^{10}\text{C}\rightarrow {}^{10}\text{B}$. It was first computed using nuclear shell model, which reported $-0.345(35)\%$~\cite{Hardy:2014qxa}. Later, the dispersive analysis unveiled important missing nuclear effects in such calculation, in particular the quasi-elastic contribution~\cite{Seng:2018qru}; such contribution was estimated with a crude Fermi gas model, which led to a substantial shift of the central value as well as an inflated uncertainty: $-0.400(50)\%$. Finally, the recent NCSM calculation reports $-0.422(31)\%$, which agrees well with the DR result while re-achieving the precision level reported by the shell model calculation. Similar calculation with NCSM is planned to be carried out for the next lightest superallowed transition, ${}^{14}\text{O}\rightarrow {}^{14}\text{N}$. 

\subsection{Nucleus-dependent RC: effective field theory approach}

\begin{figure}[tb]
	\centering
	\includegraphics[scale=0.4]{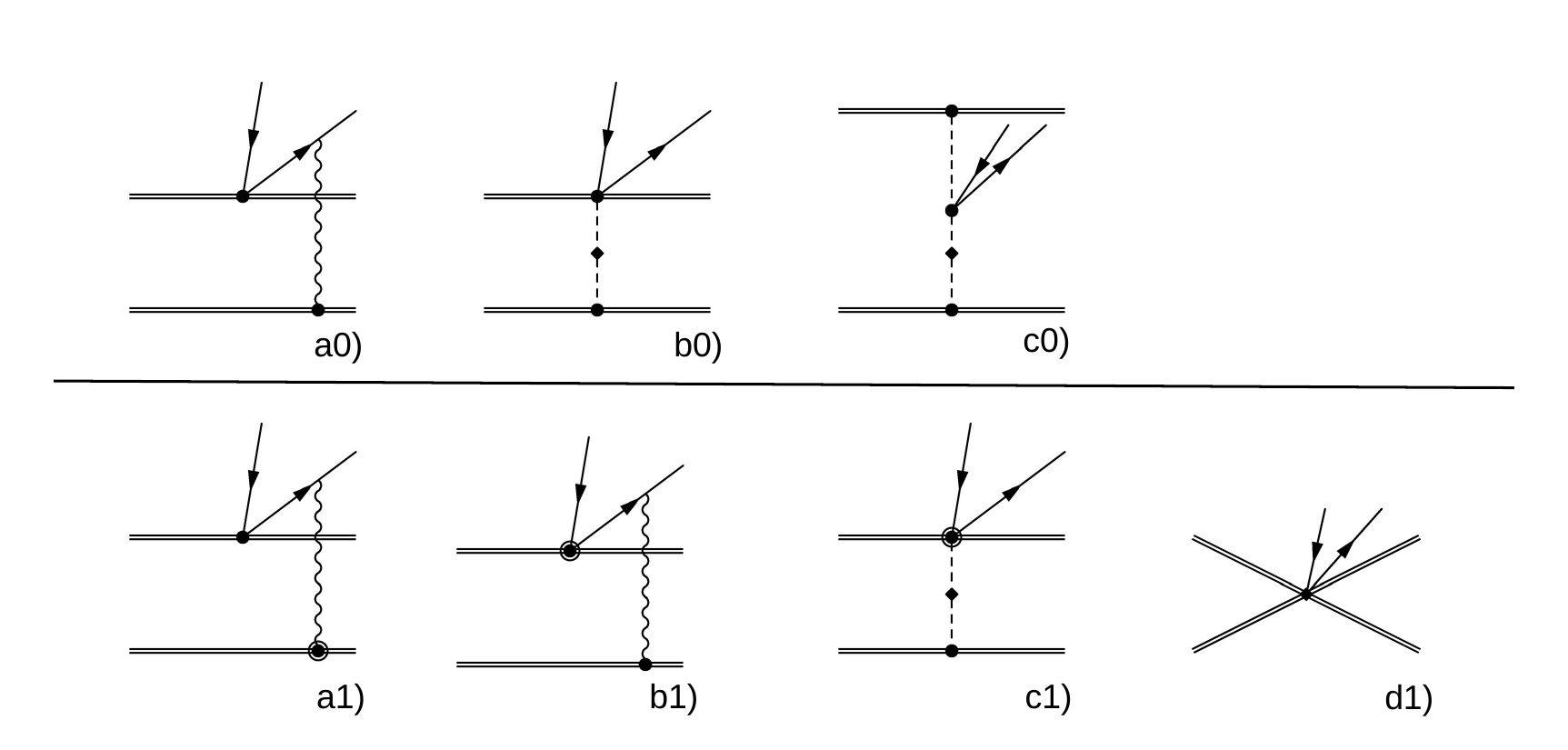}\hfill
	\caption{\label{fig:EFT} Feynman diagrams for two-body potentials in the EFT approach to $\delta_\text{NS}$. Figure taken from Ref.\cite{Cirigliano:2024msg} with permission.} 
\end{figure}

An alternative approach to $\delta_\text{NS}$ based on effective field theory (EFT) was developed in parallel to the dispersive approach~\cite{Cirigliano:2024rfk,Cirigliano:2024msg}. It consists of splitting the relevant regions of loop integral into the ``ultrasoft'' region ($q_0\sim \vec{q}\sim 10^0~\text{MeV}$) and the ``potential'' region ($q_0\sim 10^0~\text{MeV}\ll\vec{q}\sim 10^2~\text{MeV}$). It is shown that contributions from the ultrasoft region are dominated by elastic intermediate states and can be computed analytically; meanwhile, contributions from the potential region are largely independent of intermediate states, and thus can be represented as ground-state nuclear matrix elements of various effective two-body potentials, derived from Feynman diagrams in  Fig.\ref{fig:EFT}. Computing such matrix elements is numerically less demanding than computing the nuclear Green's function, as it does not requires summing up all nuclear intermediate states. 

It was shown that the nuclear matrix element of the long distance two-body potential in Fig.4(a1) contains ultraviolet divergences, which need to be reabsorbed into two low energy constants (LECs) that appear in short-distance two-body counterterm potentials depicted in Fig.4(d1). The same also happens in the EFT description of neutrinoless double beta decays~\cite{Cirigliano:2018hja}. These LECs are not fixed by chiral symmetry, and at the present can only be estimated using na\"{\i}ve dimensional analysis. They represent the major source of uncertainty in the determination of $\delta_\text{NS}$ with the EFT approach. Refs.\cite{Cirigliano:2024msg,Cirigliano:2024rfk} presented the first EFT-based \textit{ab initio} calculation of $\delta_\text{NS}$ in ${}^{14}\text{O}\rightarrow {}^{14}\text{N}$ with Quantum Monte Carlo (QMC) method. They reported $V_{ud}[^{14}\text{O}]=0.97364(52)_{\delta_\text{NS}}(20)_{\text{other}}$, which compares well with $V_{ud}[^{14}\text{O}]=0.97405(31)_{\delta_\text{NS}}(20)_{\text{other}}$ in Ref.\cite{Hardy:2020qwl} but with a larger total uncertainty. The authors proposed to perform a combined fit of $V_{ud}$ and the two LECs to all the measured rates of superallowed nuclear transitions in order to determine them simultaneously. 

\section{Summary and outlook}

In this talk I emphasized the need for highly accurate theory prediction of the SM RC for the precision test of the SM with beta decays. Theory inputs governed by non-perturbative QCD are required in this process. At the hadron level, they have recently been determined to a satisfactory level of accuracy using DR + experimental data and lattice QCD. It is highly desirable to average over the results of $\Delta_R^V$ from these two methods in the future, and to do so requires a separation between the ``Born'' and ``non-Born'' contribution to $\Delta_R^V$ in the full lattice result, which is non-trivial at the present and is a future direction of investigation. 

At the nuclear level, \textit{ab initio} nuclear many-body calculations are needed to pin-down the nucleus-structure-dependent RC $\delta_\text{NS}$. 
Two different theoretical approaches are available: The dispersive approach explicitly accounts for all nucleonic contributions to $\delta_\text{NS}$, but is computationally more demanding as it requires a summation over all nuclear intermediate states; on the other hand, the EFT approach simplifies the numerical calculation, but it is limited by the unknown LECs.

An important observation is that the physics encoded in the LECs are in fact explicitly present in the $\delta_\text{NS}$ computed with the dispersive approach. An example is the quasi-elastic nucleon contribution, which involves the single-nucleon form factors convoluted with the nucleon momentum distribution in a nucleus; such contribution, which extends to a few hundred MeV in $\vec{q}$, is not present in the explicit one-nucleon diagrams in the EFT and thus has to reside in the LECs. Therefore, an extremely powerful strategy would be to perform a DR-EFT matching: One may compute relevant components of $\delta_\text{NS}$ for a couple of superallowed transitions, with the same \textit{ab initio} method, in both the dispersive and EFT approach, and equate the two results. This gives two equations, sufficient to solve for the two unknown LECs. Once these LECs are fixed, one can simply proceed with the EFT approach for other superallowed transitions. 

To construct the matching relation, one needs to identify the terms in the EFT potentials that originate from the axial $\gamma W$-box diagram, because the latter is what forms $\delta_\text{NS}$ in the dispersive approach. These involves the components of the ``magnetic'' (mag) and ``recoil'' (rec) potential linear to $g_A$, as well as the full counterterm (CT) potential. The correct matching relation reads: 
\begin{equation}
	\delta_\text{NS}=\sqrt{2}\langle \mathcal{V}_0^\text{mag}+\mathcal{V}_0^{\text{rec,1}}+\mathcal{V}_0^{\text{CT}}\rangle_{fi}~,\label{eq:matching}
\end{equation}
which may serve as the starting point for the aforementioned matching analysis. The two lightest superallowed transitions, 
${}^{10}\text{C}\rightarrow {}^{10}\text{B}$ and ${}^{14}\text{O}\rightarrow {}^{14}\text{N}$, are possible candidates for this program, since the full $\delta_\text{NS}$ in these transitions can be computed reliably using, say, NCSM. The completion of this analysis in the near future may lead to another significant improvement in the theory prediction of $\delta_\text{NS}$, which may further refine the existing determination of $V_{ud}$ and provide stronger constrains on possible BSM physics residing in charged weak decay processes.

\section{Acknowledgment}
This work is supported in part by the U.S. Department of Energy (DOE), Office of Science, Office of Nuclear Physics, under award DE-FG02-97ER41014, by the FRIB Theory Alliance award DE-SC0013617, by the DOE Topical Collaboration "Nuclear Theory for New Physics", award No. DE-SC0023663, and by University of Tennessee, Knoxville.

\bibliography{ref}

\begin{thebibliography}{28}

\bibitem{Cabibbo:1963yz}
N.~Cabibbo, {Unitary Symmetry and Leptonic Decays}, Phys. Rev. Lett.
  \textbf{10}, 531 (1963). \doiwoc{10.1103/PhysRevLett.10.531}

\bibitem{Kobayashi:1973fv}
M.~Kobayashi, T.~Maskawa, {CP Violation in the Renormalizable Theory of Weak
  Interaction}, Prog. Theor. Phys. \textbf{49}, 652 (1973).
  \doiwoc{10.1143/PTP.49.652}

\bibitem{Hardy:2020qwl}
J.C. Hardy, I.S. Towner, {Superallowed $0^+ \to 0^+$ nuclear $\beta$ decays:
  2020 critical survey, with implications for V$_{ud}$ and CKM unitarity},
  Phys. Rev. C \textbf{102}, 045501 (2020).
  \doiwoc{10.1103/PhysRevC.102.045501}

\bibitem{Seng:2018yzq}
C.Y. Seng, M.~Gorchtein, H.H. Patel, M.J. Ramsey-Musolf, {Reduced Hadronic
  Uncertainty in the Determination of $V_{ud}$}, Phys. Rev. Lett. \textbf{121},
  241804 (2018), \texttt{1807.10197}. \doiwoc{10.1103/PhysRevLett.121.241804}

\bibitem{Seng:2018qru}
C.Y. Seng, M.~Gorchtein, M.J. Ramsey-Musolf, {Dispersive evaluation of the
  inner radiative correction in neutron and nuclear $\beta$ decay}, Phys. Rev.
  \textbf{D100}, 013001 (2019), \texttt{1812.03352}.
  \doiwoc{10.1103/PhysRevD.100.013001}

\bibitem{Gorchtein:2023srs}
M.~Gorchtein, C.Y. Seng, {The Standard Model Theory of Neutron Beta Decay},
  Universe \textbf{9}, 422 (2023), \texttt{2307.01145}.
  \doiwoc{10.3390/universe9090422}

\bibitem{Feng:2020zdc}
X.~Feng, M.~Gorchtein, L.C. Jin, P.X. Ma, C.Y. Seng, {First-principles
  calculation of electroweak box diagrams from lattice QCD}, Phys. Rev. Lett.
  \textbf{124}, 192002 (2020), \texttt{2003.09798}.
  \doiwoc{10.1103/PhysRevLett.124.192002}

\bibitem{Ma:2023kfr}
P.X. Ma, X.~Feng, M.~Gorchtein, L.C. Jin, K.F. Liu, C.Y. Seng, B.G. Wang, Z.L.
  Zhang, {Lattice QCD Calculation of Electroweak Box Contributions to
  Superallowed Nuclear and Neutron Beta Decays}, Phys. Rev. Lett. \textbf{132},
  191901 (2024), \texttt{2308.16755}. \doiwoc{10.1103/PhysRevLett.132.191901}

\bibitem{Barker:1991tw}
F.C. Barker, B.A. Brown, W.~Jaus, G.~Rasche, {Determination of V (ud) from
  Fermi decays and the unitarity of the KM mixing matrix}, Nucl. Phys. A
  \textbf{540}, 501 (1992). \doiwoc{10.1016/0375-9474(92)90171-F}

\bibitem{Towner:1992xm}
I.S. Towner, {The Nuclear structure dependence of radiative corrections in
  superallowed Fermi beta decay}, Nucl. Phys. A \textbf{540}, 478 (1992).
  \doiwoc{10.1016/0375-9474(92)90170-O}

\bibitem{Towner:1994mw}
I.S. Towner, {Quenching of spin operators in the calculation of radiative
  corrections for nuclear beta decay}, Phys. Lett. B \textbf{333}, 13 (1994),
  \texttt{nucl-th/9405031}. \doiwoc{10.1016/0370-2693(94)91000-6}

\bibitem{Towner:2002rg}
I.S. Towner, J.C. Hardy, {Calculated corrections to superallowed Fermi beta
  decay: New evaluation of the nuclear structure dependent terms}, Phys. Rev. C
  \textbf{66}, 035501 (2002), \texttt{nucl-th/0209014}.
  \doiwoc{10.1103/PhysRevC.66.035501}

\bibitem{Towner:2007np}
I.S. Towner, J.C. Hardy, {An Improved calculation of the
  isospin-symmetry-breaking corrections to superallowed Fermi beta decay},
  Phys. Rev. C \textbf{77}, 025501 (2008), \texttt{0710.3181}.
  \doiwoc{10.1103/PhysRevC.77.025501}

\bibitem{Seng:2022cnq}
C.Y. Seng, M.~Gorchtein, {Dispersive formalism for the nuclear structure
  correction \ensuremath{\delta}NS to the \ensuremath{\beta} decay rate}, Phys.
  Rev. C \textbf{107}, 035503 (2023), \texttt{2211.10214}.
  \doiwoc{10.1103/PhysRevC.107.035503}

\bibitem{Gorchtein:2023naa}
M.~Gorchtein, C.Y. Seng, {Superallowed nuclear beta decays and precision tests
  of the Standard Model}, Ann. Rev. Nucl. Part. Sci. \textbf{74}, 23 (2024),
  \texttt{2311.00044}. \doiwoc{10.1146/annurev-nucl-102622-020726}

\bibitem{Gennari:2024sbn}
M.~Gennari, M.~Drissi, M.~Gorchtein, P.~Navratil, C.Y. Seng, {Ab~Initio
  Strategy for Taming the Nuclear-Structure Dependence of Vud Extractions: The
  C10{\textrightarrow}B10 Superallowed Transition}, Phys. Rev. Lett.
  \textbf{134}, 012501 (2025), \texttt{2405.19281}.
  \doiwoc{10.1103/PhysRevLett.134.012501}

\bibitem{Barrett:2013nh}
B.R. Barrett, P.~Navratil, J.P. Vary, {Ab initio no core shell model}, Prog.
  Part. Nucl. Phys. \textbf{69}, 131 (2013).
  \doiwoc{10.1016/j.ppnp.2012.10.003}

\bibitem{haydock1974inverse}
R.~Haydock, The inverse of a linear operator, Journal of Physics A:
  Mathematical, Nuclear and General \textbf{7}, 2120 (1974).

\bibitem{Dagotto:1993ajt}
E.~Dagotto, {Correlated electrons in high-temperature superconductors}, Rev.
  Mod. Phys. \textbf{66}, 763 (1994), \texttt{cond-mat/9311013}.
  \doiwoc{10.1103/RevModPhys.66.763}

\bibitem{Marchisio:2002jx}
M.A. Marchisio, N.~Barnea, W.~Leidemann, G.~Orlandini, {Lorentz integral
  transform for inclusive and exclusive cross-sections with the Lanczos
  method}, Few Body Syst. \textbf{33}, 259 (2003), \texttt{nucl-th/0202009}.
  \doiwoc{10.1007/s00601-003-0017-z}

\bibitem{Donnelly:1975ze}
T.W. Donnelly, J.D. Walecka, {Electron Scattering and Nuclear Structure}, Ann.
  Rev. Nucl. Part. Sci. \textbf{25}, 329 (1975).
  \doiwoc{10.1146/annurev.ns.25.120175.001553}

\bibitem{Donnelly:2017aaa}
T.W. Donnelly, J.A. Formaggio, B.R. Holstein, R.G. Milner, B.~Surrow,
  {Foundations of Nuclear and Particle Physics} (Cambridge University Press,
  2017), ISBN 978-1-108-11018-1, 978-0-521-76511-4

\bibitem{glauber1959}
R.J. Glauber, {High energy collision theory}, Lectures in theoretical physics,
  Interscience, New York London  (1959).

\bibitem{Gribov:1968jf}
V.N. Gribov, {Glauber corrections and the interaction between high-energy
  hadrons and nuclei}, Sov. Phys. JETP \textbf{29}, 483 (1969).

\bibitem{Hardy:2014qxa}
J.C. Hardy, I.S. Towner, {Superallowed $0^+\to 0^+$ nuclear \ensuremath{\beta}
  decays: 2014 critical survey, with precise results for $V_{ud}$ and CKM
  unitarity}, Phys. Rev. C \textbf{91}, 025501 (2015), \texttt{1411.5987}.
  \doiwoc{10.1103/PhysRevC.91.025501}

\bibitem{Cirigliano:2024msg}
V.~Cirigliano, W.~Dekens, J.~de~Vries, S.~Gandolfi, M.~Hoferichter,
  E.~Mereghetti, {Ab initio electroweak corrections to superallowed
  {\ensuremath{\beta}} decays and their impact on Vud}, Phys. Rev. C
  \textbf{110}, 055502 (2024), \texttt{2405.18464}.
  \doiwoc{10.1103/PhysRevC.110.055502}

\bibitem{Cirigliano:2024rfk}
V.~Cirigliano, W.~Dekens, J.~de~Vries, S.~Gandolfi, M.~Hoferichter,
  E.~Mereghetti, {Radiative Corrections to Superallowed {\ensuremath{\beta}}
  Decays in Effective Field Theory}, Phys. Rev. Lett. \textbf{133}, 211801
  (2024), \texttt{2405.18469}. \doiwoc{10.1103/PhysRevLett.133.211801}

\bibitem{Cirigliano:2018hja}
V.~Cirigliano, W.~Dekens, J.~De~Vries, M.L. Graesser, E.~Mereghetti,
  S.~Pastore, U.~Van~Kolck, {New Leading Contribution to Neutrinoless
  Double-{\ensuremath{\beta}} Decay}, Phys. Rev. Lett. \textbf{120}, 202001
  (2018), \texttt{1802.10097}. \doiwoc{10.1103/PhysRevLett.120.202001}

\end{thebibliography}

\end{document}